%% file: main.tex
\documentclass{article}
\usepackage{arxiv}
\usepackage[utf8]{inputenc} 
\usepackage[T1]{fontenc}    
\usepackage{hyperref}       
\usepackage{url}            
\usepackage{booktabs}       
\usepackage{amsfonts}       
\usepackage{nicefrac}       
\usepackage{microtype}      
\usepackage{lipsum}		
\usepackage{graphicx}
\usepackage{natbib}
\usepackage{doi}
\usepackage{braket}
\usepackage{float}
\usepackage[textsize=small]{todonotes}
 \usepackage{amsmath}
 \usepackage{subcaption}

\title{Option Pricing on Noisy Intermediate-Scale Quantum Computers: A Quantum Neural Network Approach}

\author{Sebastian Zaj\c{a}c\\
	finQbit\\
	Warsaw, Poland \\
	\texttt{sebastian.zajac@finqbit.tech} \\
	\And
	Rafał Pracht \\
	finQbit\\
	Warsaw, Poland\\
	\texttt{rafal.pracht@finqbit.tech} \\
}

\renewcommand{\shorttitle}{QNN for Option Pricing on NISQ Devices}

\extrafloats{100}
\begin{document}
\maketitle

\begin{abstract}
In a global derivatives market with notional values in the hundreds of trillions of dollars, the accuracy and efficiency of pricing models are of fundamental importance, with direct implications for risk management, capital allocation, and regulatory compliance. In this work, we employ the Black-Scholes-Merton (BSM) framework not as an end in itself, but as a controlled benchmark environment in which to rigorously assess the capabilities of quantum machine learning methods.

We propose a fully quantum approach to option pricing based on Quantum Neural Networks (QNNs), and, to the best of our knowledge, present one of the first implementations of such a methodology on currently available quantum hardware. Specifically, we investigate whether QNNs, by exploiting the geometric structure of Hilbert space, can effectively approximate option pricing functions.

Our implementation utilizes a compact 2-qubit QNN architecture evaluated across multiple state-of-the-art quantum processors, including IBM Fez, IQM Garnet, IonQ Forte, and Rigetti Ankaa-3. This cross-platform study reveals distinct hardware-dependent performance characteristics while demonstrating that accurate pricing approximations can be achieved consistently across different devices despite the constraints of Noisy Intermediate-Scale Quantum (NISQ) hardware.

The results provide empirical evidence that QNN-based approaches constitute a viable framework for derivative pricing. While the analysis is conducted within the BSM setting, the broader significance lies in the potential extension of these methods to more realistic and computationally demanding models, including local volatility, stochastic volatility, and interest rate frameworks commonly used in practice.

\end{abstract}
\input{01_intro}
\input{02_background_classic}
\input{02_background_qnn}
\input{03_Methodology}
\input{04_experiments}

\input{05_results}

\bibliographystyle{unsrtnat}
\bibliography{references}  
\end{document}

%% file: 01_intro.tex
\section{Introduction}

Derivative pricing and option valuation, in particular, constitute central components of modern financial systems. 
Financial institutions such as banks, asset managers, and insurance companies rely extensively on derivative instruments for risk management, portfolio optimization, and hedging strategies. 
According to the Bank for International Settlements (BIS) \cite{BIS2023OTC}, the global over-the-counter (OTC) derivatives market consistently exhibits notional values on the order of hundreds of trillions of U.S. dollars. 
A substantial portion of this market is characterized by embedded optionality, which appears either explicitly through listed and OTC options or implicitly within structured products and credit instruments.

Given the scale of this market, even marginal improvements in pricing accuracy or computational efficiency can translate into significant economic impact, including capital savings and enhanced risk control. 
From an operational standpoint, financial institutions perform millions of valuations on a daily basis for trading, risk management, regulatory reporting, and stress testing purposes. 
As financial instruments become increasingly complex and regulatory requirements under the Basel III framework become more demanding, there is a growing need for computational methods that are both accurate and scalable while maintaining operational efficiency.

To manage the immense volume and complexity of these instruments, the industry requires standardized mathematical frameworks. 
In this context, the Black–Scholes–Merton (BSM) framework \cite{BlackScholes1973, Merton1973} remains a foundational model in derivative pricing, owing to its closed-form analytical solutions for European-style options, which provide both computational efficiency and conceptual clarity. 
In practice, derivative valuations are frequently expressed in terms of implied volatility rather than price, reflecting the central role of the BSM model as a reference framework. 
Beyond its direct applicability, the BSM model serves as a conceptual and mathematical benchmark for more advanced approaches, including local volatility models \cite{DermanSmile, Dupire1994PricingWA}, stochastic volatility models \cite{Heston1993, Hagan2002SABR}, and jump-diffusion extensions \cite{Merton1976Jump}, as well as evolving term structures, including modern interest rate frameworks like LIBOR Market Model \cite{Brace1997Market}.

However, the elegance of closed-form solutions is lost when moving beyond basic European options. 
Many financial products exhibit path-dependent payoffs or feature early exercise rights as seen in American-style options, often involving multiple underlying assets. 
In such high-dimensional settings, analytical solutions are unavailable, necessitating a reliance on numerical techniques. 
While Monte Carlo simulations are flexible, their convergence is slow, and finite difference methods suffer from the curse of dimensionality, where computational costs grow exponentially with the number of risk factors. 
It is precisely this bottleneck in scaling that motivates the exploration of Quantum Machine Learning (QML) and Quantum Monte Carlo (QMC), which may offer more efficient ways to represent and process these complex, multi-dimensional payoffs. 
Quantum Monte Carlo methods have been previously investigated in \cite{PrachtBinomialTree}; however, their practical implementation requires more advanced quantum hardware than is currently available. 
Consequently, in this work, we focus on QML approaches.

The objective of this work is to investigate whether QML techniques \cite{Schuld2019Hilbert,Schuld2021Kernel,Schuld2024Inductive}, and in particular Quantum Neural Networks \cite{Farhi2018QNN,Abbas_2021}, can effectively approximate option prices within the Black-Scholes-Merton framework. 
The use of the BSM model as a testbed is intentional. 
While BSM does not suffer from the computational bottlenecks of higher-dimensional models, its analytical solution provides a reliable benchmark for evaluating approximation accuracy while offering a controlled environment in which to assess the capabilities and limitations of quantum learning models \cite{Herman2023QuantumFinance}.

Importantly, this study does not aim to replace classical valuation methods for standard instruments. 
Rather, it seeks to establish a proof-of-concept framework for the application of quantum machine learning to financial problems. 
The ultimate goal is to extend these methods to more complex and business-critical settings such as high-dimensional derivatives, path-dependent payoffs, and models where classical approaches encounter computational bottlenecks.

Our results demonstrate that Quantum Neural Networks can approximate BSM option prices with high precision, even when implemented on NISQ hardware \cite{Preskill2018NISQ}. 
This finding represents an important step toward the practical application of quantum technologies in derivative pricing and financial risk management, where the potential for quantum advantage is expected to be most impactful \cite{Arute2019QuantumSupremacy}.

\section{Classical Machine Learning Approaches to Option Pricing}

The convergence of high-performance computing and the ubiquity of granular financial data has catalyzed a paradigmatic shift toward machine learning (ML) within quantitative finance. 
Beyond mere computational speed, ML frameworks offer a robust alternative to classical asset pricing, particularly in circumventing rigid structural assumptions and the curse of dimensionality that often hinder traditional numerical schemes in complex option valuation \cite{Hutchinson1994, RufWang2020}.

The adoption of ML within empirical asset pricing, as surveyed by Weigand \cite{Weigand2019}, leverages the expansion of computational power to navigate previously intractable high-dimensional feature spaces. 
Unlike traditional econometric frameworks, ML architectures autonomously distill nonlinear risk factors and execute sparse feature selection through advanced regularization techniques, offering a more granular lens on market dynamics. 
Empirical evidence, particularly from deep learning and ensemble tree methods, suggests that these models yield superior predictive power for expected returns by effectively penalizing noise through regularization \cite{Gu2020}. 
Nevertheless, the inherent opacity of these estimators persists as a critical hurdle, especially regarding out-of-sample stability, interpretability, and the difficulty of establishing causal inference in non-stationary financial environments.


Neural networks parameterize option pricing functions by reformulating valuation as a supervised learning task, effectively mapping model inputs directly to contract prices. 
Recent benchmarks by Della Corte et al. \cite{dellacorte2026machinelearningoptionpricing} provide a systematic decomposition of architectural influence across both BSM and Heston environments.
By pitting feedforward and residual networks against specialized topologies such as the Deep Galerkin Method (DGM) and highway networks, they demonstrate that structural choices are non-trivial for minimizing approximation error. 
Notably, generalized highway networks exhibit superior convergence compared to standard multilayer perceptrons, while DGM-based variants excel in representing implied volatility surfaces after feature space transformations. 
The primary operational advantage lies in the train once, infer always paradigm: once offline optimization is complete, the network bypasses the latency of Monte Carlo or PDE solvers, enabling real-time valuation even for complex, high-dimensional payoffs.


The application of machine learning to American option pricing represents a significant advancement, given the inherent computational complexity associated with optimal stopping problems. Unlike European derivatives, American-style options require determining an optimal exercise strategy, which introduces a free boundary problem typically addressed through dynamic programming or iterative numerical schemes. 

Anderson and Ulrych \cite{AmericanOptionDNN} propose a deep neural network framework designed to accelerate American option pricing within a market-making context. Their methodology involves generating synthetic training data via conventional numerical solvers under the Heston stochastic volatility model and subsequently training a neural network to approximate the pricing function across the parameter space.

A primary advantage of this architecture is the strategic shift of the computational burden from online execution to offline training. Once optimized, the neural network provides nearly instantaneous price evaluations, ensuring real-time responsiveness in trading environments where minimizing latency is critical. The authors demonstrate that this approach achieves a superior balance between speed and accuracy compared to traditional techniques while maintaining the precision required for practical applications.

Furthermore, the framework emphasizes regulatory compliance and interpretability by preserving the underlying structural model. By ensuring that the neural network operates as a transparent function approximator rather than a purely data-driven black box, this methodology aligns with the growing requirement for explainable artificial intelligence (XAI) in financial systems \cite{Arrieta2020XAI}.


The classical machine learning literature provides strong evidence that neural networks can effectively approximate pricing functions for a wide range of financial instruments. 
By learning the mapping between model parameters and option prices, these methods circumvent the need for repeated numerical simulations during inference.

If classical models can navigate these complex landscapes, the question arises whether Quantum Neural Networks, leveraging quantum-enhanced feature maps and the vast representational capacity of Hilbert spaces, can offer superior or more resource-efficient approximations. Building on these classical benchmarks, this work probes the viability of quantum-enhanced surrogates, using established machine learning results as a baseline for measuring quantum utility.

%% file: 02_background_classic.tex
\begin{figure}[h!]
    \centering
\includegraphics[width=0.6\textwidth]{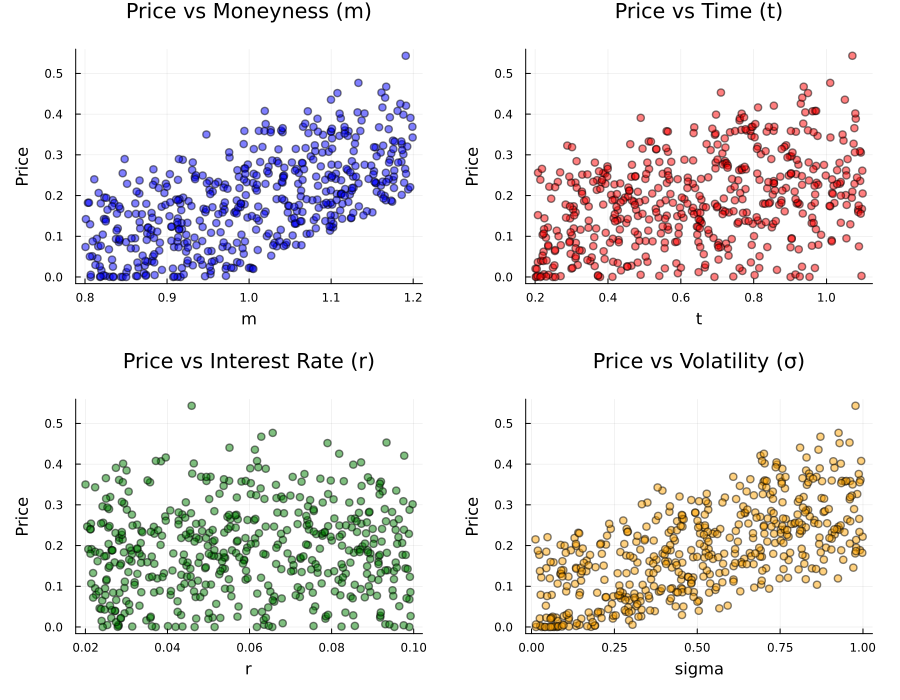}
    \caption{Scatterplots illustrating the dependency of the option price on each input parameter.}
    \label{fig:generator_scatter}
\end{figure}

\section{Experimental Methodology and Data Framework}\label{sec:data}

To benchmark the proposed quantum architecture, we constructed a synthetic data generation framework predicated on the Black-Scholes-Merton (BSM) model for European call options. This approach ensures a controlled environment where reference values are available through analytical solutions, allowing for a granular decomposition of approximation errors across distinct market regimes. The dataset comprises a feature matrix $\mathbf{X} \in \mathbb{R}^{4 \times n}$ and a corresponding target vector $\mathbf{y} \in \mathbb{R}^n$ representing option prices. 
To ensure broad coverage of market regimes, input features are independently sampled from a joint uniform distribution $\mathcal{U}(L, U)$ spanning the following dimensions:

\begin{itemize}
    \item \textbf{Moneyness} ($m = S/K$): Reflects the intrinsic value state of the option across $m \in [0.8, 1.2]$.
    \item \textbf{Time to Maturity} ($T$): Annualized residual life sampled from $T \in [0.2, 1.1]$.
    \item \textbf{Risk free Interest Rate} ($r$): Annualized discount factor where $r \in [0.02, 0.1]$.
    \item \textbf{Volatility} ($\sigma$): Annualized standard deviation of log returns across $\sigma \in [0.01, 1.0]$.
\end{itemize}

For each sample $i$, the target label $y_i$ is defined as the normalized European call price $\hat{C}_i = C_i / K$. By adopting the strike price as a numeraire, we express the analytic Black Scholes Merton solution in terms of moneyness $m = S/K$:

\begin{equation}
\hat{C} = m N(d_1) - e^{-rT} N(d_2)
\end{equation}

where $N(\cdot)$ denotes the cumulative distribution function (CDF) of the standard normal distribution. Under this parameterization, the kernels $d_1$ and $d_2$ are reformulated as follows:

\begin{equation}
d_1 = \frac{\ln(m) + \left(r + \frac{\sigma^2}{2}\right) T}{\sigma\sqrt{T}}, \quad d_2 = d_1 - \sigma\sqrt{T}
\end{equation}

This normalization is particularly advantageous for variational training. By condensing the price $S$ and strike $K$ into a single relative metric $m$, we effectively reduce the dimensionality of the input space. 
From a quantum perspective, this dimensionality reduction is critical as it allows for a more efficient encoding of features into a limited number of qubits, ensuring that the high-dimensional Hilbert space is utilized to capture genuine non-linear sensitivities rather than redundant scale information. This not only simplifies the learning landscape for the Quantum Neural Network but also ensures that the model captures the scale-invariant nature of option pricing while preserving all underlying risk sensitivities.

To ensure the robustness of the dataset, we generated a training set of $N_{\text{train}} = 500$ observations and a test set of $N_{\text{test}} = 100$\footnote{To validate the robustness of the reported results further, we conducted an additional out-of-sample evaluation using an independently generated test set of $N_{\text{test}} = 10,000$ observations. The resulting performance metrics were consistent with those obtained on the original test set, confirming that the conclusions are not sensitive to the choice of sample size and that the model generalizes reliably across larger datasets.}. 
The statistical integrity of the synthetic data was validated through Pearson correlation analysis, confirming the expected dependencies within the BSM framework. While initial training and testing were conducted on a high-fidelity simulator using the full dataset, final model performance was validated on physical quantum hardware using five distinct benchmark points. This step was critical to assess the impact of quantum noise and decoherence on pricing accuracy in a real-world execution environment.


Figure \ref{fig:generator_scatter} illustrates the distribution of generated option prices relative to each input dimension. 
The scatterplots confirm that the generator effectively populates the predefined parameter space without localized clustering, ensuring that the model is exposed to a diverse range of market conditions spanning from deep out of the money to deep in the money regimes. 
Furthermore, the clearly defined nonlinear trends, particularly with respect to volatility ($\sigma$) and moneyness ($m$), verify that the synthetic data accurately captures the characteristic curvature of the BSM pricing surface.

\subsection{Classical Benchmarking and the ATM Convexity Challenge}

To benchmark the proposed quantum architecture against established classical standards, we implemented two regression models representing opposite ends of the complexity spectrum:
\begin{enumerate}
\item \textbf{Ordinary Least Squares (OLS):} A linear estimator (5 parameters) employed to quantify the "Linearity Gap" defined here as the residual error arising from the Black Scholes surface’s deviation from a first-order hyperplane. 
This baseline isolates the nonlinear curvature that any advanced model must capture.
\item \textbf{XGBoost Regressor:} A Gradient Boosted Decision Tree \cite{Chen2016XGBoost} ensemble serving as a high capacity non linear benchmark. 
The model was configured with a learning rate ($\eta$) of $0.2$, a maximum depth of $3$, and $50$ boosting rounds. Given the constrained training set ($N_{\text{train}} = 500$), we enforced regularization through stochastic sub-sampling ($70\%$) and column sampling ($80\%$) per tree to ensure robust generalization and prevent overfitting to the synthetic manifold.
\end{enumerate}

A granular residual analysis across moneyness groups highlights a specific bottleneck we define as the \textit{ATM Convexity Challenge}. 
In derivative pricing, the At The Money (ATM) region (defined here as $0.95 \le m \le 1.05$) coincides with peak Gamma ($\Gamma = \partial^2 \hat{C} / \partial m^2$) \cite{Hull2021Options,Gatheral2011VolatilitySmile}. 
This represents the maximum curvature (convexity) of the pricing surface, where the option's sensitivity to the underlying price is most volatile. 
Capturing this sharp non-linearity is a primary hurdle for any estimator, as it marks the transition point where the option's extrinsic value is most concentrated.

Our analysis reveals that while XGBoost performs admirably in the tails, its approximation error significantly spikes within the ATM interval. As shown in Table \ref{tab:classical_granular_baseline}, the MSE for XGBoost in the ATM region ($0.000328$) nearly doubles compared to the OTM and ITM regimes.

\begin{table}[h!]
\centering
\caption{Granular MSE Analysis of Classical Baselines.}
\label{tab:classical_granular_baseline}
\begin{tabular}{lcc}
\toprule
\textbf{Moneyness Regime} & \textbf{OLS (MSE)} & \textbf{XGBoost (MSE)} \\ \midrule
OTM ($m < 0.95$)           & 0.000658           & 0.000171               \\
ATM ($0.95 \le m \le 1.05$) & 0.000618           & 0.000328               \\
ITM ($m > 1.05$)           & 0.000771           & 0.000183               \\ \bottomrule
\end{tabular}
\end{table}
This performance dip stems from the inherent limitations of tree-based ensembles: their reliance on axis-aligned, piecewise constant approximations makes it difficult to smoothly interpolate the rapid, sigmoidal transition of the option's Delta.
This bottleneck underscores the potential for architectures like the proposed QNN.
By leveraging the continuous geometry of Hilbert spaces, QNNs may natively capture high-dimensional nonlinearities without the need for the excessive parameterization often required by classical surrogates. 
To establish a rigorous baseline, we evaluate OLS and XGBoost on an out-of-sample test set ($N=100$), pinpointing the specific regimes where classical heuristics falter. 
The global performance metrics for these baseline models are summarized in Table \ref{tab:classical_results_global}.

\begin{table}[h!]
\centering
\caption{Global performance comparison of classical baseline models.}
\label{tab:classical_results_global}
\begin{tabular}{lcccc}
\toprule
\textbf{Model} & \textbf{MSE} & \textbf{RMSE} & \textbf{MAE} & \textbf{$R^2$} \\ \midrule
OLS (Linear Baseline) & 0.00069 & 0.02618 & 0.01971 & 0.93291 \\
XGBoost (Non linear)  & 0.00022 & 0.01480 & 0.01215 & 0.97854 \\ \bottomrule
\end{tabular}
\end{table}

%% file: 02_background_qnn.tex
\section{Quantum Neural Architectures for Derivative Pricing}

Traditional numerical methods, such as Monte Carlo simulations or finite difference schemes, while accurate, become computationally expensive as the dimensionality of the problem increases.
QNNs offer an alternative paradigm where non-linear financial dependencies are mapped directly into a high-dimensional Hilbert space \cite{Abbas_2021}. 

While the majority of QML literature focuses on classification tasks, the application of QNNs for continuous function approximation \cite{Mitarai_2018} and financial time-series analysis \cite{Cybulski2024} remains a developing field. 
A critical challenge in modeling the Black-Scholes (BSM) pricing surface is capturing the high convexity in the At-The-Money region. 
This requires models with high expressivity that remain hardware-efficient, minimizing the gate count to mitigate noise in the NISQ era.

\subsection{4-Qubit Quantum Neural Networks}

The initial phase of our research focused on a classical variational approach where each input feature $\mathbf{x} = [m, T, r, \sigma]^\top$ (see. Chapter \ref{sec:data}) is mapped to a dedicated qubit. 
This model utilizes the expectation value of the $\hat{Z}$ operator on the readout qubit.
Given the nature of financial instruments, where the price $\hat{C} \ge 0$, a max function is applied as a post-processing step:
\begin{equation}
    \hat{C}_{\text{pred}} = \max(0,  \bra{0} U^{\dagger}(\theta,x)\, \hat{Z} \, U(\theta,x) \ket{0})
\end{equation}

This approach ensures that physical constraints (non-negative pricing) are embedded into the estimator structure, enhancing training stability against hardware-induced residuals by filtering out non-physical fluctuations during the optimization process.

\begin{table}[h!]
\centering
\caption{Baseline QNN model performance on test set (4-qubit register).}
\label{tab:qnn_baseline}
\begin{tabular}{ccccccc}
\toprule
\textbf{Layers ($L$)} & \textbf{Parameters} & \textbf{CX Gates} & \textbf{MSE} & \textbf{RMSE} & \textbf{MAE} & \textbf{$R^2$} \\ \midrule
1 & 12 & 4  & 0.00641 & 0.08004 & 0.06816 & 0.53004 \\
2 & 24 & 8  & 0.00042 & 0.02037 & 0.01544 & 0.96955 \\
3 & 36 & 12 & 0.00030 & 0.01738 & 0.01172 & \textbf{0.97783} \\
4 & 48 & 16 & 0.00030 & 0.01726 & 0.01206 & 0.97814 \\ \bottomrule
\end{tabular}
\end{table}

Based on empirical findings, the 3-layer configuration emerges as the most optimal solution for this architecture. 
While further training may marginally improve convergence, the 3-layer model provides a superior balance between high fidelity approximation ($R^2 \approx 0.978$) and resource parsimony. 

\subsection{The Fourier Perspective and Data Re-uploading}

Following the framework established by \cite{Schuld2021}, QNNs can be interpreted as partial Fourier series where the number of data encoding stages (\textit{Data Re-uploading}) determines the available frequency spectrum. 
Models defined as \cite{P_rez_Salinas_2020}:
\begin{equation}
f_{\theta}(x) = \bra{0} U^{\dagger}(x,\theta)\, \hat{Z} \, U(x, \theta) \ket{0}
\end{equation} 
where $U(x,\theta)$ is a quantum circuit that depends on the input $x$ and parameters $\theta$, can be represented as a Fourier series: 
\begin{equation}
f_{\theta}(x) = \sum_{\omega \in \Omega} c_{\omega}(\theta)e^{i \omega x}
\end{equation}

Circuit model $U(x,\theta)$ can be written as: 
\begin{equation}
U^{L}(x, \theta) = W^{L+1}(\theta)\, S(x) \,W^{L}(\theta) \dots W^2(\theta)\,S(x)\, W^1(\theta)
\end{equation}
The frequency spectrum $\Omega$ is determined by the eigenvalues of the encoding generators in $S(x)$. 
As demonstrated by Schuld et al. \cite{Schuld2021}, increasing $L$ linearly expands the available frequencies in the model's Fourier representation. 
This expansion is crucial for capturing the high-frequency components of the Black-Scholes surface, allowing the model to resolve higher-order derivatives of the pricing surface.

To test this hypothesis, we conducted experiments on a 4-qubit register, varying the re-uploading depth $L$. 
In this architecture, an initial variational layer $W_0$ is applied before the first data injection to provide an optimized starting state in the Hilbert space. 
Consequently, a circuit with $L$ layers actually consists of $L+1$ trainable weight blocks, providing additional degrees of freedom for the optimization process.
\begin{table}[h!]
\centering
\caption{Regression performance for the Fourier-enhanced 4-qubit model (including $W_0$).}
\label{tab:qnn_fourier}
\begin{tabular}{ccccccc}
\toprule
\textbf{Layers ($L$)} & \textbf{Parameters} & \textbf{CX Gates} & \textbf{MSE} & \textbf{RMSE} & \textbf{MAE} & \textbf{$R^2$} \\ \midrule
1 & 24 & 8  & 0.00639 & 0.07991 & 0.06789 & 0.53151 \\
2 & 36 & 12 & 0.04070 & 0.20173 & 0.16451 & -1.98551 \\
3 & 48 & 16 & 0.00034 & 0.01853 & 0.01351 & \textbf{0.97481} \\
4 & 60 & 20 & 0.00021 & 0.01446 & 0.00907 & 0.98466 \\ \bottomrule
\end{tabular}
\end{table}

The failure of the $L=2$ model to generalize, despite a near-perfect training fit, suggests that the optimization landscape for 36 parameters across 12 CX gates may be prone to local minima or Barren Plateaus \cite{McClean2018}, where the circuit's inductive bias is misaligned with the BSM manifold.

As with the baseline model, the $L=3$ configuration remains the most optimal solution for practical deployment, balancing precision with a manageable resource overhead.

\subsection{The finQbit Architecture: Compression and Feature Shuffling}
To address the resource intensity of 4-qubit models and the limitations of NISQ hardware, we propose the \textbf{finQbit} architecture. This model reduces the quantum register to an $n=2$ qubit register, maintaining -- and in certain regimes exceeding -- the precision of higher-qubit alternatives through two fundamental innovations:
\begin{enumerate}
    \item \textbf{High-Density Encoding:} Each qubit encodes two features simultaneously using orthogonal rotation axes ($R_y, R_x$), maximizing the information density per qubit.
    \item \textbf{Dynamic Feature Permutation:} Feature pairs mapped to each qubit are permuted across the 3-layer re-uploading scheme to capture diverse inter-feature correlations.
\end{enumerate}

The circuit $U(\mathbf{x}, \theta, \phi)$ consists of $L=3$ layers of alternating trainable entangling blocks $W^{(k)}$ and data-encoding blocks $S^{(k)}$. The output is derived from the expectation value of the $\hat{Z}$ operator on the readout qubit:
\begin{equation}
\hat{C}_{\text{pred}} = \max(0, \bra{0} U^{\dagger}(\mathbf{x}, \theta, \phi) \,\, \hat{Z}_0 \,\,U(\mathbf{x}, \theta, \phi) \ket{0})
\end{equation}

\subsection{Variational Layers ($W$)}
Each variational block $W^{(k)}$ applies universal $U3$ rotations to all qubits, followed by a bidirectional entanglement sequence:
\begin{equation}
W^{(k)} = \text{CNOT}_{1,0} \cdot \text{CNOT}_{0,1} \cdot \left( U3(\vec{\theta}_{0}^{\,k}) \otimes U3(\vec{\theta}_{1}^{\,k}) \right)
\end{equation}
where $\vec{\theta}_{q}^{\,k} = (\theta_{q,1}^k, \theta_{q,2}^k, \theta_{q,3}^k)$ denotes the vector of trainable Euler angles for qubit $q$ in layer $k$. This sequence ensures that the variational state is fully entangled before each data injection stage.

\subsection{Permutation-Based Data Re-uploading ($S$)}
To overcome the limitations of the reduced qubit count, we implement a dynamic encoding scheme $S^{(k)}(\mathbf{x}, \phi^{(k)})$. 
In each layer $k \in \{1, 2, 3\}$, the input features $(m, \sigma, T, r)$ are mapped to the rotation angles of $R_x$ and $R_y$ gates, scaled by trainable parameters $\phi^{(k)}$\footnote{In contrast to established Python-based frameworks like PennyLane or Qiskit, which typically enforce a rigid separation between data features and variational weights (often requiring complex \textit{ParameterExpression} binding or custom templates), \textbf{finQbit} allows for native, integrated learnable data scaling directly within the encoding layers.}. 
These features were formally defined in Chapter \ref{sec:data}. 
Crucially, the mapping is permuted in each stage:
\begin{itemize}
    \item \textbf{Layer $S^{(1)}$}: Maps $(m, \sigma) \to Q_0$ and $(T, r) \to Q_1$.
    \item \textbf{Layer $S^{(2)}$}: Permutes the input to $(T, m) \to Q_0$ and $(r, \sigma) \to Q_1$.
    \item \textbf{Layer $S^{(3)}$}: Reconfigures the pairs to $(m, T) \to Q_0$ and $(r, \sigma) \to Q_1$.
\end{itemize}
The general form of the encoding operator for a qubit $q$ in layer $k$ is defined as:
\begin{equation}
S_q^{(k)}(x_i, x_j) = R_y(q, \phi_{j}^{k} \cdot x_j) \cdot R_x(q, \phi_{i}^{k} \cdot x_i)
\end{equation}

\subsection{Resource Complexity}
The finQbit architecture is optimized for parameter efficiency and low circuit depth. The total count of 36 trainable parameters is distributed between variational weights and encoding scalers as shown in Table \ref{tab:parameter_distribution}.

\begin{table}[h]
\centering
\caption{Distribution of trainable parameters in the finQbit architecture ($L=3$).}
\label{tab:parameter_distribution}
\begin{tabular}{llcc}
\toprule
\textbf{Component} & \textbf{Description} & \textbf{Calculation} & \textbf{Count} \\ \midrule
Variational Weights ($\theta$) & $U3$ angles in $L+1$ blocks & $4 \text{ blocks} \times 6$ & 24 \\
Encoding Scalers ($\phi$) & Feature weights in $L$ blocks & $3 \text{ blocks} \times 4$ & 12 \\ \midrule
\textbf{Total Parameters} & & & \textbf{36} \\ \bottomrule
\end{tabular}
\end{table}

\begin{figure}[h]
    \centering
    \includegraphics[width=0.8\linewidth]{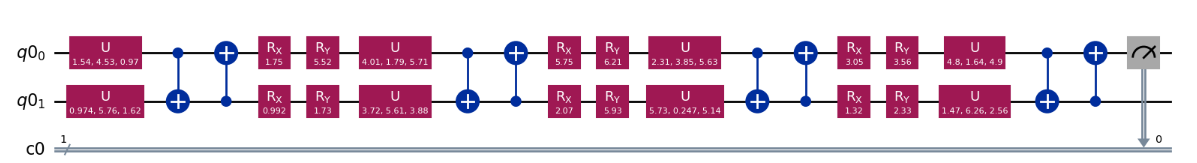}
    \caption{The finQbit VQC architecture illustrating the 2-qubit register.}
    \label{fig:circuit_design}
\end{figure}

The finQbit model was evaluated against classical benchmarks, including OLS and XGBoost. 
As summarized in Table \ref{tab:final_results_global}, the proposed quantum model achieved an $R^2$ of $0.97958$, slightly surpassing the performance of XGBoost. 
This suggests that Hilbert space rotations provide a superior inductive bias for capturing the continuous, smooth nature of the Black-Scholes pricing manifold compared to the discrete decision boundaries of tree-based models.

\begin{table}[h!]
\centering
\caption{Global performance comparison on the out-of-sample test set ($N=100$).}
\label{tab:final_results_global}
\begin{tabular}{lcccc}
\toprule
\textbf{Model} & \textbf{MSE} & \textbf{RMSE} & \textbf{MAE} & \textbf{$R^2$} \\ \midrule
OLS (Linear Baseline) & 0.00069 & 0.02618 & 0.01971 & 0.93291 \\
XGBoost (Non-linear)  & 0.00022 & 0.01480 & 0.01215 & 0.97854 \\
\textbf{finQbit} & \textbf{0.00021} & \textbf{0.01444} & \textbf{0.01081} & \textbf{0.97958} \\ \bottomrule
\end{tabular}
\end{table}


Table \ref{tab:final_granular_comparison} shows the MSE across different moneyness regimes. 
Notably, in the ATM region ($0.95 \le m \le 1.05$)
the finQbit model reduced the MSE by approximately 50\% compared to XGBoost. This confirms that continuous unitary transformations are more adept at approximating high-convexity surfaces than piecewise-constant approximations.

\begin{table}[h]
\centering
\caption{Granular MSE Analysis across Moneyness Regimes.}
\label{tab:final_granular_comparison}
\begin{tabular}{lccc}
\toprule
\textbf{Moneyness Regime} & \textbf{OLS (MSE)} & \textbf{XGBoost (MSE)} & \textbf{QNN (MSE)} \\ \midrule
OTM ($m < 0.95$)          & 0.000658           & 0.000171               & \textbf{0.000074} \\
ATM ($0.95 \le m \le 1.05$) & 0.000618           & 0.000328               & \textbf{0.000163} \\
ITM ($m > 1.05$)          & 0.000771           & \textbf{0.000183}      & 0.000212 \\ \bottomrule
\end{tabular}
\end{table}

Furthermore, the superior predictive power of the finQbit model is achieved with remarkable structural parsimony. 
As summarized in Table \ref{tab:complexity}, the quantum architecture requires only 36 trainable parameters, an order of magnitude reduction compared to the approximately 400 parameters utilized by the optimized XGBoost ensemble.

\begin{table}[ht!]
\centering
\caption{Comparison of model complexity and parameter count.}
\label{tab:complexity}
\begin{tabular}{lcl}
\toprule
\textbf{Model} & \textbf{Parameter Count} & \textbf{Approximation Type} \\ \midrule
OLS            & 5                        & Linear                      \\
XGBoost        & $\sim$400                & Piecewise-constant (Trees)   \\
\textbf{finQbit} & \textbf{36}              & Continuous Unitary (Hilbert) \\ \bottomrule
\end{tabular}
\end{table}

\subsection{Ansatz Optimization and Unitary Decomposition} \label{KAK_Decomposition}

To mitigate the adverse effects of hardware noise, the original QNN architecture was systematically compressed into a minimal, hardware-efficient representation. 
Specifically, the initial multi-gate structure, comprising eight controlled-NOT gates, was reduced to a single 2-qubit unitary block in $U(4)$ via state vector-based optimization. 
This transformation significantly decreases circuit depth and the accumulation of incoherent errors, thereby enhancing the fidelity of state preparation on NISQ devices.

\begin{figure}[h]
    \centering
    \includegraphics[width=0.5\linewidth]{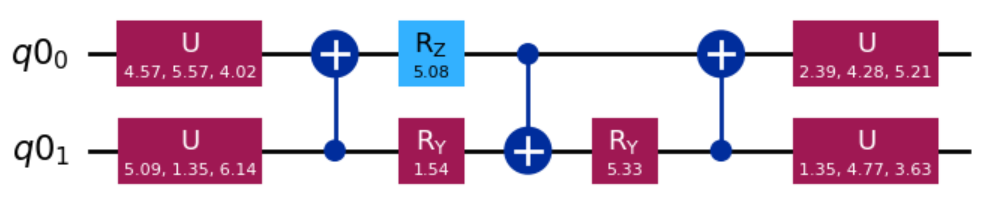}
    \caption{The finQbit VQC architecture after U4 decomposition gate.}
    \label{fig:circuit_design}
\end{figure}

The parametrization of the resulting $U(4)$ unitary was obtained using a machine learning–based optimization procedure. 
We define a loss function as the matrix norm between the target unitary, corresponding to the original QNN circuit, and the unitary generated by the parametrized $U(4)$ block. 
Formally, the objective is to minimize the discrepancy between these two operators over the parameter space of the ansatz. 
The optimization was performed using standard gradient-based methods, yielding a set of parameters for which the resulting unitary closely approximates the target transformation.

This compression strategy enables a substantial reduction in circuit complexity while preserving the predictive performance of the model, as measured by the coefficient of determination ($R^2$). 
Consequently, it provides an effective pathway for deploying expressive QNN architectures within the constraints of current quantum hardware.

%% file: 03_Methodology.tex
\section{Simulation Execution Analysis}

To assess the empirical performance and practical viability of the \textit{finQbit} model under realistic operating conditions, the quantum circuits were transpiled into formats compatible with both the AWS Braket SDK and IBM Qiskit frameworks. 
The benchmarking methodology is organized into a three-tier hierarchy designed to systematically isolate distinct sources of error and to optimize the allocation of computational resources:
\begin{enumerate}
\item \textbf{Ideal State Vector Simulation}, performed using the \textit{finQbit} simulator, to establish a theoretical baseline in the absence of sampling noise and hardware imperfections;
\item \textbf{Shot Noise Simulation}, conducted via AWS Braket and Qiskit simulators, to quantify statistical sampling errors arising from finite measurement counts\footnote{Stemming from the collapse of the wavefunction $|\psi\rangle$ during the measurement process, where the estimator variance is asymptotically bounded by $1/\sqrt{N_{\text{shots}}}$}, while excluding hardware-induced decoherence;
\item \textbf{Execution on Physical QPUs}, across multiple hardware backends including IQM, IonQ, Rigetti, and IBM, to evaluate the combined effects of gate infidelity, decoherence, and device-specific noise characteristics.
\end{enumerate}

Given the substantial operational costs and queue latencies associated with access to QPUs, a detailed sensitivity analysis was performed to identify an optimal trade-off between statistical precision and economic efficiency. 
The experimental design explores a range of sampling configurations defined by shot counts $N_{\text{shots}} \in \{500, 2000, 5000\}$ and repetition counts $R \in \{20, 50\}$. 
This analysis is essential for determining the point of diminishing returns, at which further reductions in estimator variance no longer justify the incremental cost of additional quantum executions.

To evaluate the robustness of the model across representative market conditions, a controlled parameter space was employed. 
Specifically, the time-to-maturity $T = 1.0$, risk-free interest rate $r = 0.05$, and volatility $\sigma = 0.2$ were held constant, thereby isolating the model’s sensitivity to the \textbf{moneyness} parameter ($m$), which serves as the primary determinant of the curvature of the BSM pricing surface. 
The model was benchmarked across five discrete moneyness levels spanning the full spectrum from in-the-money to out-of-the-money regimes: $m \in \{0.8, \,\, 0.9, \,\, 1.0, \,\, 1.1, \,\, 1.2\}$.

\subsection{Ideal Simulation}

The \textit{finQbit} framework was initially validated using state-vector simulations in the Julia environment. This step was crucial to establish an idealized analytical baseline, isolating the model's intrinsic expressive power from the stochastic sampling noise (shot noise) and hardware-induced decoherence typical of real quantum processors. 
\begin{figure}[h]
    \centering
    \includegraphics[width=0.6\linewidth]{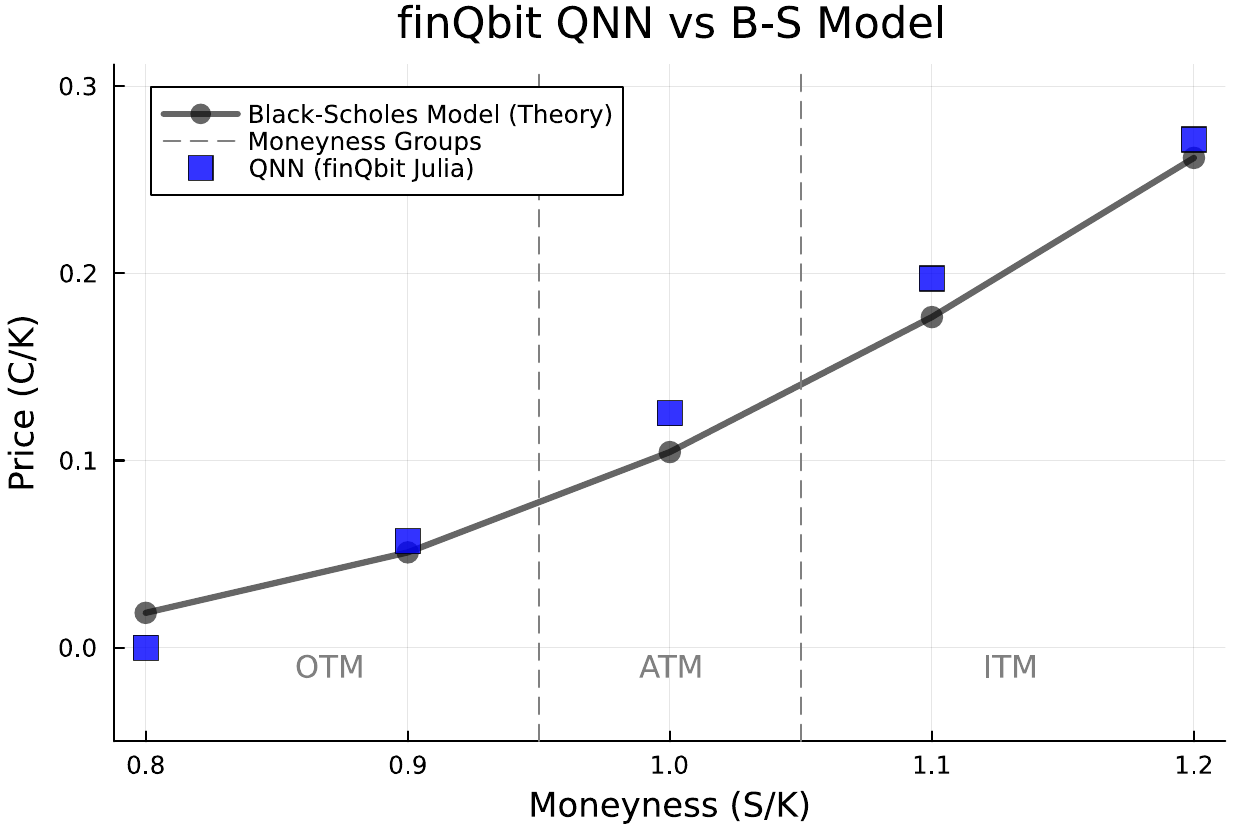}
    \caption{Comparison between the finQbit predictions and the analytical Black-Scholes prices across the tested moneyness spectrum.}
    \label{fig:julia_plot}
\end{figure}

As illustrated in Figure \ref{fig:julia_plot}, the discrete QNN predictions exhibit a high degree of alignment with the continuous Black-Scholes price curve. 
Each of the five representative market configurations is reconstructed, demonstrating that the model captures the non-linear profile of the call option price. 

The statistical synthesis of this simulation is presented in Table \ref{tab:julia_eval}. 
\begin{table}[H]
\centering
\caption{Performance metrics for the finQbit model across the 5-point test set.}
\label{tab:julia_eval}
\begin{tabular}{llc}
\toprule
\textbf{Metric} & \textbf{Full Name} & \textbf{Value} \\ \midrule
$R^2$ & Coefficient of Determination & \textbf{0.96537} \\
MSE & Mean Squared Error & 0.00027 \\
RMSE & Root Mean Squared Error & 0.01634 \\
MAE & Mean Absolute Error & 0.01516 \\
Max Error & Maximum Prediction Error & 0.02075 \\ \bottomrule
\end{tabular}
\end{table}
The high $R^2$ value (\textbf{0.96537}) and a remarkably low Mean Squared Error (MSE) confirm that the 2-qubit \textit{finQbit} architecture, despite its structural parsimony, effectively approximates the target financial transformation. These results serve as a high-fidelity reference for the subsequent analysis of hardware constraints within the AWS Braket and IBM Quantum ecosystems.

\subsection{Shot Noise Simulation}

To bridge the gap between idealized state vector simulations and realistic execution environments, we conduct experiments using count-based simulators available within \textbf{AWS Braket} and \textbf{IBM Qiskit}. 
The primary objective of this analysis is to determine the \textbf{statistical convergence threshold}, defined as the minimum number of measurement samples required to achieve financial-grade accuracy while maintaining cost-efficient utilization of quantum processing resources.

The empirical results obtained using \textbf{AWS Braket} are summarized in Table~\ref{tab:stats_combined}. 
These results exhibit a clear manifestation of the Law of Large Numbers: as the number of shots $N_{\text{shots}}$ increases, both the standard deviation ($\sigma$) and the maximum error decrease, leading to progressively more stable and accurate predictions.

\begin{table}[H]
\centering
\caption{QNN Experiment Statistics for $R=20$ and $R=50$ Repetitions (AWS Braket).}
\label{tab:stats_combined}
\begin{tabular}{@{}lcccc@{}}
\toprule
\textbf{Config ($R$, Shots)} & \textbf{MAE} & \textbf{Std. Dev ($\sigma$)} & \textbf{Max Error} & \textbf{$R^2$} \\ \midrule
$R=20, N=500$  & 0.0213 & 0.0414 & 0.0448 & 0.912 \\
$R=20, N=2000$ & 0.0191 & 0.0221 & 0.0281 & 0.948 \\
\textbf{R$=20$, N$=5000$} & \textbf{0.0195} & \textbf{0.0133} & \textbf{0.0283} & \textbf{0.961} \\ \midrule
$R=50, N=500$  & 0.0155 & 0.0464 & 0.0263 & 0.940 \\
\textbf{R$=50$, N$=2000$} & \textbf{0.0178} & \textbf{0.0229} & \textbf{0.0253} & \textbf{0.966} \\
$R=50, N=5000$ & 0.0152 & 0.0148 & 0.0241 & 0.965 \\ \bottomrule
\end{tabular}
\end{table}

To ensure that the observed behavior is not platform-specific, an identical experimental protocol was implemented using the \textbf{IBM Qiskit AerSimulator}. 
A comparison of the AWS Braket results with those obtained from IBM Aer (Table~\ref{tab:qiskit_results}) reveals a high degree of consistency across all performance metrics. 

\begin{table}[H]
\centering
\caption{QNN Experiment Statistics for $R=20$ and $R=50$ Repetitions (IBM Qiskit).}
\label{tab:qiskit_results}
\begin{tabular}{@{}lcccc@{}}
\toprule
\textbf{Config ($R$, Shots)} & \textbf{MAE} & \textbf{Std. Dev ($\sigma$)} & \textbf{Max Error} & \textbf{$R^2$} \\ \midrule
$R=20, N=500$  & 0.0165 & 0.0432 & 0.0261 & 0.958 \\
$R=20, N=2000$ & 0.0157 & 0.0216 & 0.0251 & 0.961 \\
$R=20, N=5000$ & 0.0150 & 0.0121 & 0.0237 & 0.966 \\ \midrule
$R=50, N=500$  & 0.0140 & 0.0408 & 0.0247 & 0.966 \\
\textbf{R$=50$, N$=2000$} & \textbf{0.0128} & \textbf{0.0195} & \textbf{0.0212} & \textbf{0.974} \\
$R=50, N=5000$ & 0.0146 & 0.0111 & 0.0211 & 0.968 \\ \bottomrule
\end{tabular}
\end{table}

The transition from deterministic state vector simulations to stochastic shot-based sampling provides several important insights into the convergence properties and noise resilience of the model:
\begin{itemize}
    \item \textbf{Shot Noise Suppression and Error Scaling:} Increasing the number of measurements from $N_{\text{shots}} = 500$ to $2000$ yields an approximate 37\% reduction in the maximum error. This behavior is consistent with the theoretical $1/\sqrt{N_{\text{shots}}}$ convergence rate, confirming that statistical averaging effectively mitigates measurement-induced fluctuations.

    \item \textbf{Sampling Budget Optimization:} The results reveal a clear trade-off between the number of repetitions ($R$) and the number of shots per circuit ($N_{\text{shots}}$). 
    An optimal configuration is achieved at $R = 50$ and $N_{\text{shots}} = 2000$, which provides a favorable balance between accuracy and computational cost. 
    In contrast, when the number of repetitions is limited (e.g., $R = 20$), a higher shot count ($N_{\text{shots}} = 5000$) is required to maintain comparable predictive performance.
\end{itemize}

These findings also highlight a fundamental limitation inherent to quantum measurement processes, which can be interpreted as sampling from a Bernoulli distribution. 
Accurate estimation of the probability amplitude associated with the $\ket{1}$ state necessitates a sufficiently large number of measurements to control statistical variance. 
This limitation motivates the exploration of more advanced quantum estimation techniques. 
In particular, Quantum Amplitude Estimation (QAE)–type algorithms, as discussed in \cite{PrachtAmerican}, provide a principled framework for reducing sampling complexity while preserving estimation accuracy. 
The integration of such methods represents a promising direction for future research, with the potential to substantially enhance the efficiency of quantum-based financial modeling.

%% file: 04_experiments.tex
\section{Cross-Platform Performance on Physical Quantum Hardware}

To assess the portability and robustness of the finQbit model, we conducted a cross-platform execution across four distinct Quantum Processing Units. 
This diverse selection allows for a direct comparison between different qubit technologies, specifically superconducting transmon qubits (IQM, Rigetti, IBM) and trapped ion systems (IonQ). 
Each backend presents a unique noise profile, gate fidelity, and native gate set, providing a rigorous testbed for the model's resilience to hardware-induced decoherence.

\input{04_IQM}

\input{04_IONQ}

\input{04_Rigetti}

\input{04_IBM}

%% file: 04_IQM.tex
\subsection{Experiments on the IQM Garnet Processor}
\begin{figure}[H]
    \centering
\includegraphics[width=0.7\linewidth]{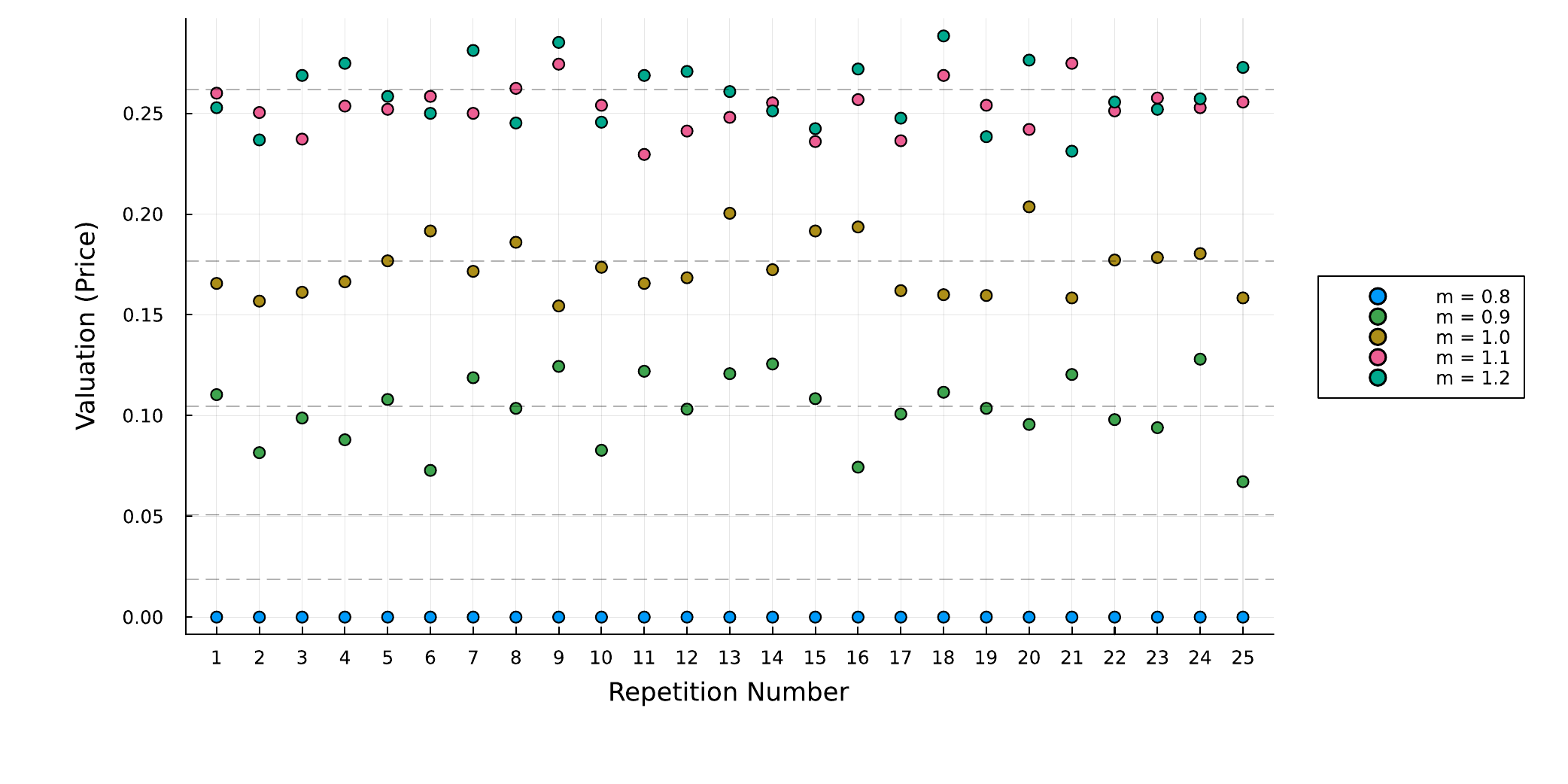}
    \caption{Repetition Stability on IQM Garnet ($R=25$). 
    The dashed lines represent Black-Scholes theoretical values.}
    \label{fig:iqm_stability}
\end{figure}

The hardware validation phase commenced with the IQM Garnet processor, a 20-qubit superconducting system accessed via the AWS Braket cloud \cite{iqm2025garnet}. As a leading example of European quantum technology, IQM Garnet provides a high-fidelity environment suitable for complex QNN executions \cite{oralkhan2026}. To assess the system's resilience to temporal noise and parameter drift, a \textit{Stability Track} experiment was conducted, consisting of $R=25$ repetitions with $5000$ shots each (see Figure \ref{fig:iqm_stability}). 
The results indicate high system consistency throughout the experimental window.

\begin{figure}[H]
    \centering
\includegraphics[width=0.5\linewidth]{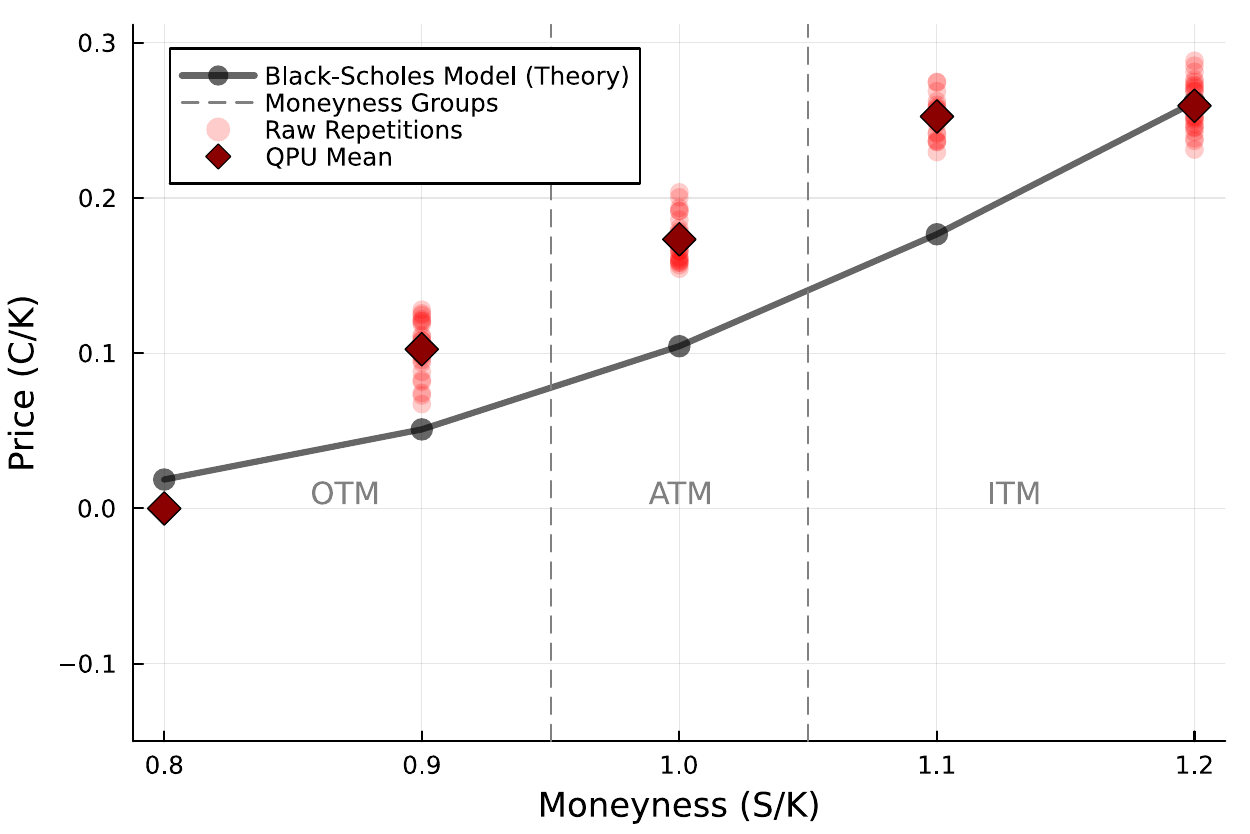}
    \caption{Point cloud distribution for IQM Garnet.}
    \label{fig:iqm_cloud}
\end{figure}

The distribution of valuation samples across different moneyness levels is shown in Figure \ref{fig:iqm_cloud}. The tight clustering of points indicates high operational repeatability. 
While a readout error mitigation protocol was applied to the results, it resulted in a further upward shift of the price estimates, maintaining the characteristic overestimation profile relative to the Black-Scholes baseline.

\begin{figure}[H]
    \centering
    \includegraphics[width=1.0\linewidth]{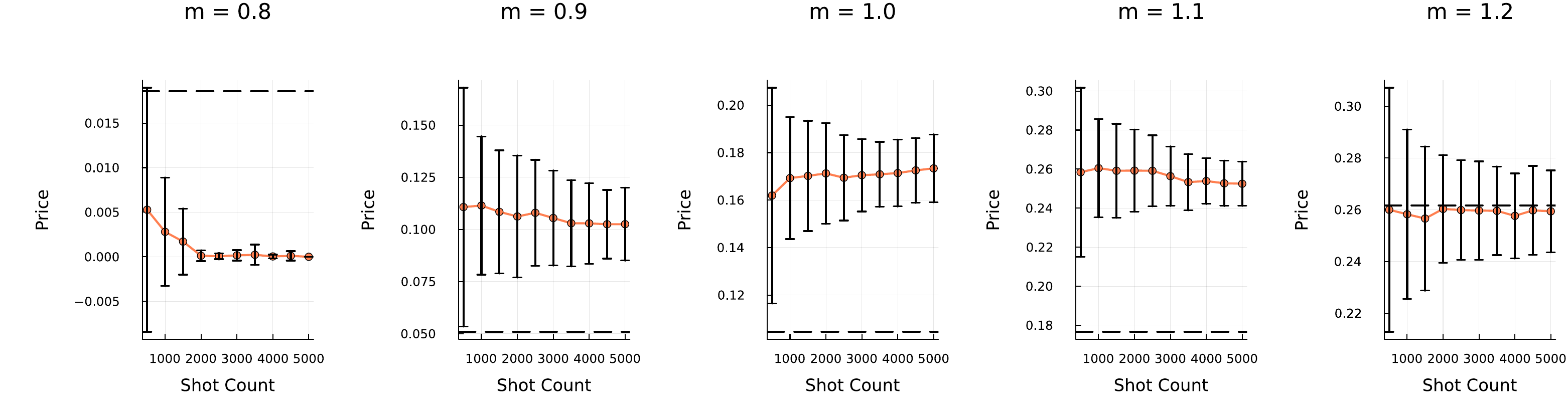}
    \caption{Convergence of the option price estimator on IQM Garnet across 500 to 5,000 shots.}
    \label{fig:iqm_convergence}
\end{figure}

To determine the optimal sampling depth for financial accuracy, a convergence analysis was performed on IQM Garnet, varying the shot count from 500 to 5,000. This study tracks the stabilization of the price estimator and the reduction of statistical noise (Fig. \ref{fig:iqm_convergence}).

The analysis of the IQM Garnet results reveals statistical stability within the 3,000–5,000 shot range, where mean valuation fluctuations settle below 1.5\% following $1/\sqrt{N}$ scaling. 
A non-linear systematic bias is observed: the processor consistently overestimates prices in the At-the-Money region ($M=1.0$, QPU $\approx 0.173$ vs. BS $\approx 0.104$), yet achieves high convergence with theoretical values at Deep In-the-Money levels ($M=1.2$, $0.259$ vs. $0.261$).
Conversely, for Out-of-the-Money options ($M=0.8$), the increasing shot count reveals a systematic downward bias that pushes the raw estimates below zero. These values are subsequently anchored to a zero valuation by the post-processing $max(x, 0)$ function, effectively masking the theoretical signal as it falls below the hardware’s operational noise floor.

%% file: 04_IONQ.tex
\subsection{Experiments on the  IonQ Forte}

IonQ Forte employs trapped-ion technology, utilizing 35 $^{171}Yb^{+}$ ions in a linear Paul trap manipulated by precision-steered lasers \cite{ionq2025aria}. 
This architecture exhibits a noise profile distinct from superconducting systems, characterized by high thermal resilience and sensitivity to laser frequency variations. 

\begin{figure}[H]
    \centering
    \includegraphics[width=0.7\linewidth]{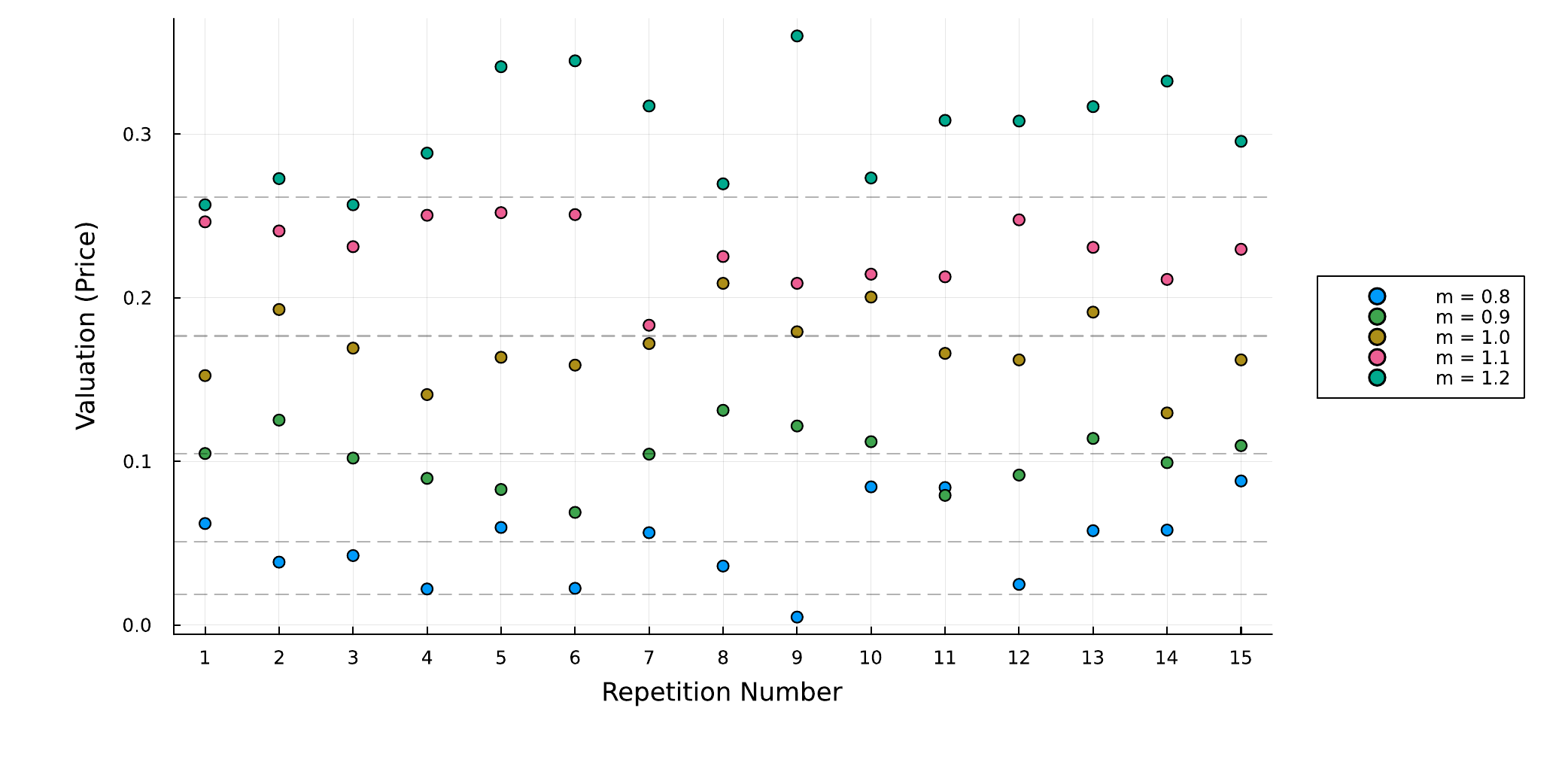}
    \caption{Drift analysis on IonQ Forte ($R=15$).}
    \label{fig:ionq_drift}
\end{figure}
The \textit{Stability Track} (Figure \ref{fig:ionq_drift}) ($R=15, 5,000$ shots) demonstrates exceptional operational consistency, maintaining a stable performance baseline with minimal variance across the experimental duration.
The choice of $R=15$ repetitions reflects a calculated balance between statistical significance and the practical constraints of high-fidelity QPU resource allocation.
\begin{figure}[H]
    \centering
    \includegraphics[width=0.5\linewidth]{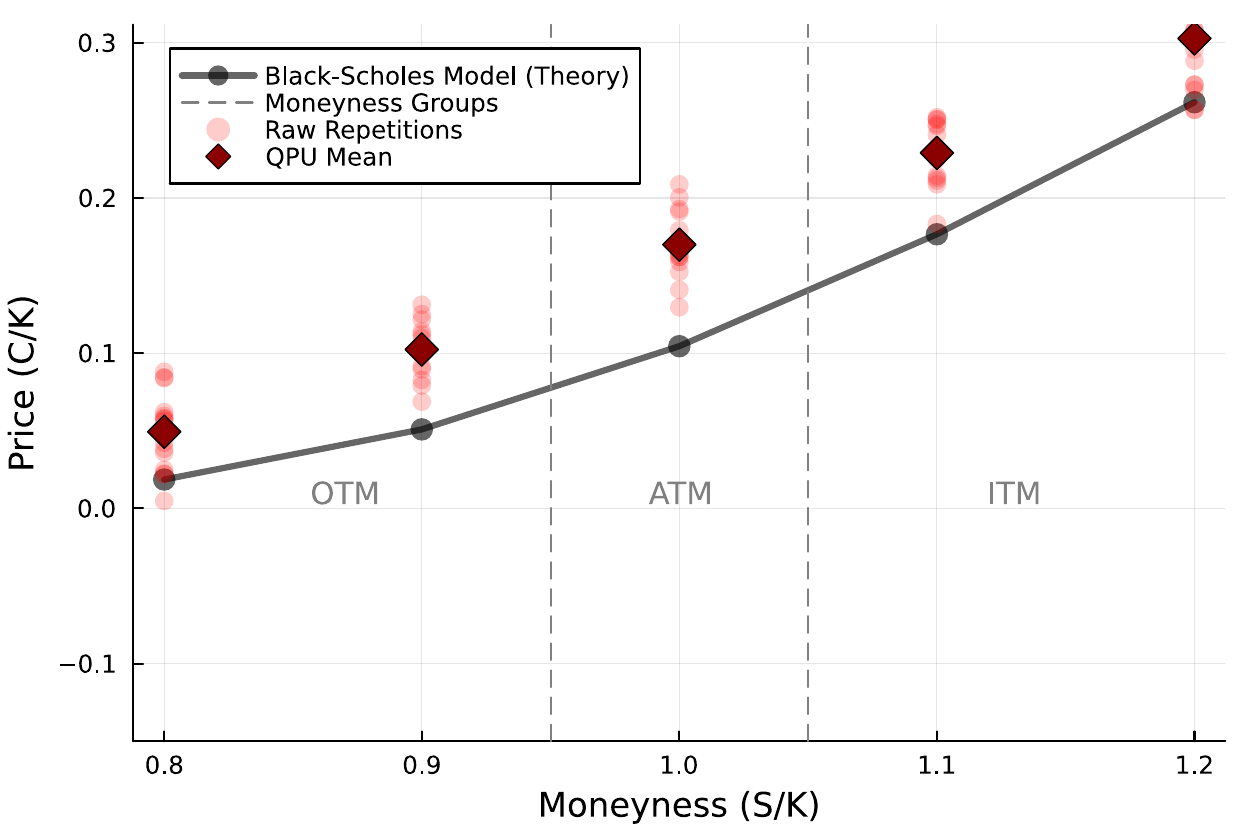}
    \caption{Point cloud distribution on IonQ Forte.}
    \label{fig:ionq_cloud}
\end{figure}

The distribution of individual valuation samples is further detailed in the point cloud analysis (Figure \ref{fig:ionq_cloud}). The results exhibit a high degree of clustering, confirming that the systematic upward bias is consistent across all moneyness levels. Notably, in the Out-of-the-Money region ($M=0.8$), the architecture maintains a positive signal-to-noise ratio, with estimates fluctuating between $0.004$ and $0.088$. Unlike superconducting systems, the IonQ Forte does not suffer from zero-price collapse at the payoff boundary, demonstrating superior signal preservation despite the persistent positive offset.

\begin{figure}[H]\centering
\includegraphics[width=1.0\linewidth]{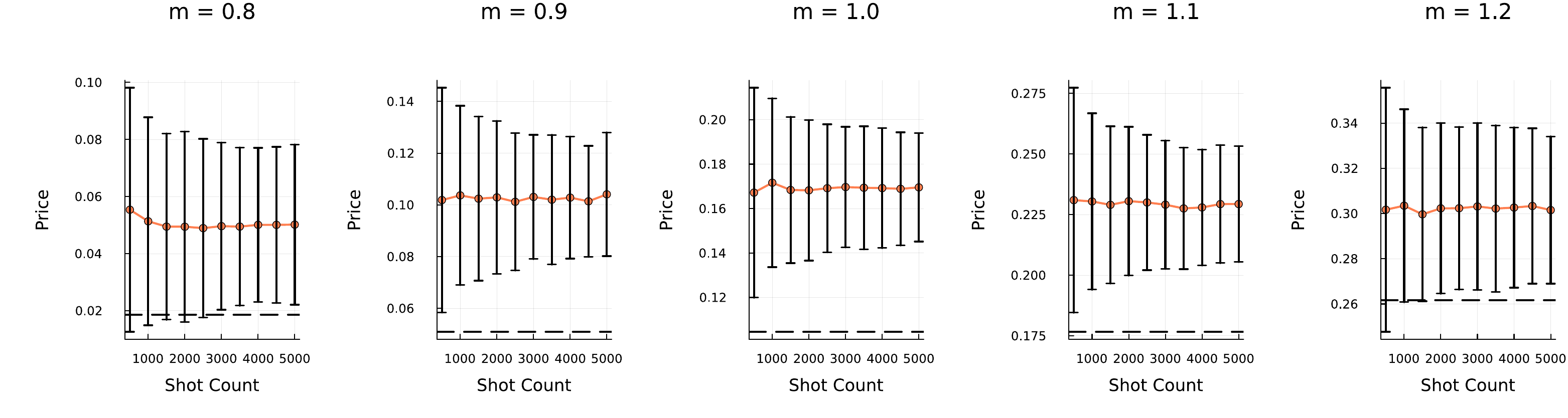}
\caption{Option price convergence trajectory for IonQ Forte. 
Values reconstructed via Monte Carlo sampling across various Moneyness levels.}
\label{fig:ionq_convergence}
\end{figure}

The Monte Carlo reconstruction of the IonQ Forte data demonstrates (Figure \ref{fig:ionq_convergence}) rapid statistical stabilization, with the standard error decreasing monotonically and reaching a plateau at approximately 3,000 shots. Beyond this threshold, mean price fluctuations remain below 0.5\%, identifying the 2,000–3,000 shot range as the optimal balance between precision and resource cost for this architecture. 
A consistent systematic overestimation is observed across the entire spectrum; for instance, at $M=1.0$, the stabilized mean of $\approx 0.1695$ exhibits a persistent gap relative to the theoretical baseline of $\approx 0.1045$. 
Notably, the estimator maintains high stability even at the $M=0.8$ level; unlike superconducting systems, the IonQ results do not collapse toward the zero floor, preserving a clear signal even for Out-of-the-Money configurations.

%% file: 04_Rigetti.tex
\subsection{Experiments on the Rigetti Ankaa-3}
Rigetti Ankaa-3 \cite{rigetti2025systems} is a late-generation superconducting QPU featuring a square lattice topology of transmon qubits. 
The Ankaa architecture focuses on utilizing tunable couplers to achieve high-fidelity multi-qubit operations, which is critical for the execution of parametrized QNN circuits.

\begin{figure}[h]
    \centering
    \includegraphics[width=0.6\linewidth]{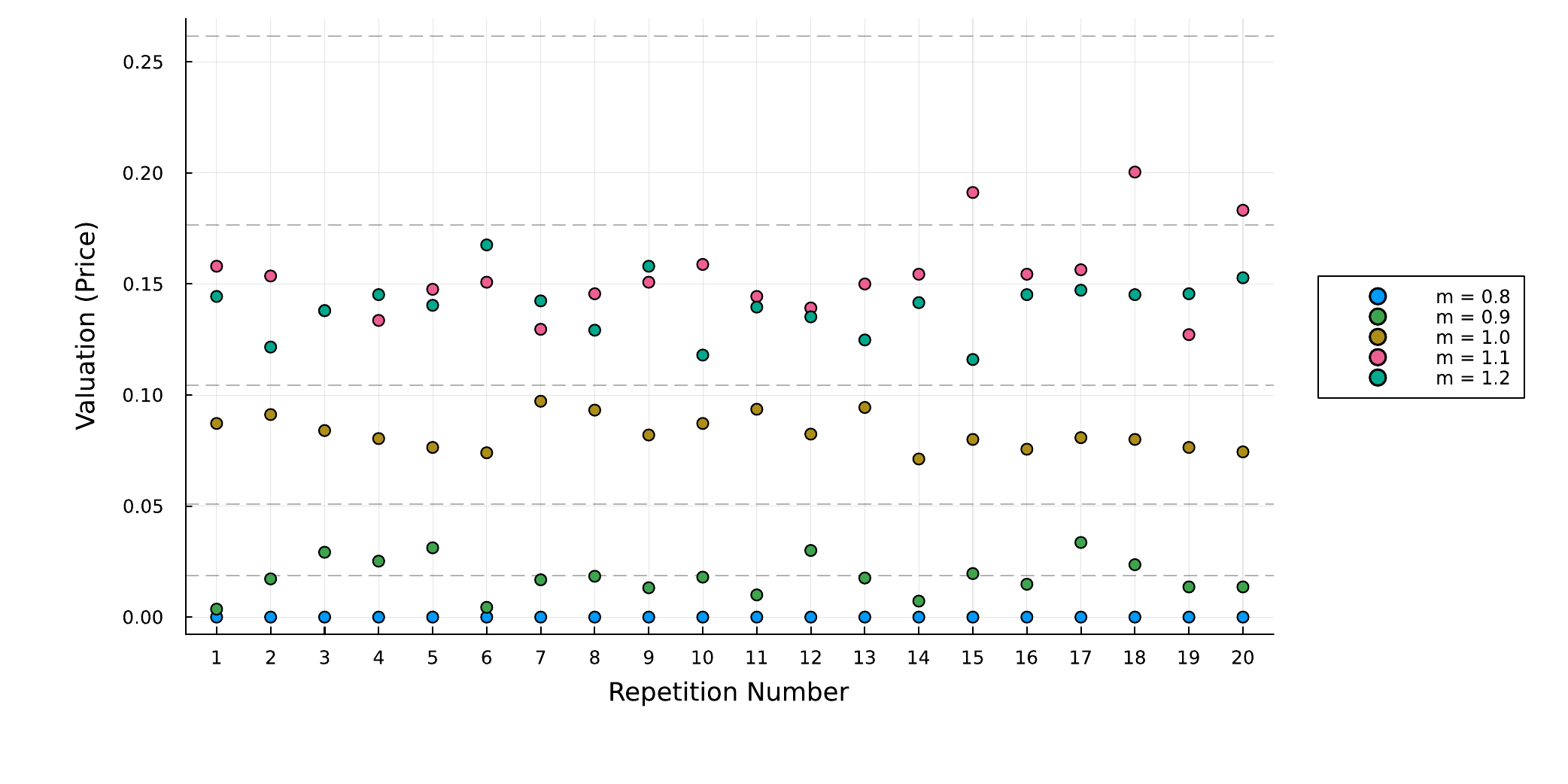}
    \caption{Drift analysis on Rigetti Ankaa-3 ($R=20$).}
    \label{fig:rigetti_drift}
\end{figure}

The stability analysis ($R=20$) for the Rigetti Ankaa-3 system (Fig. \ref{fig:rigetti_drift}) reveals a noise profile characteristic of superconducting transmons. 
The results indicate a measurable temporal drift in the price estimates, particularly for In-the-Money (ITM) configurations. 
While the Out-of-the-Money ($M=0.8$) valuations remain consistently anchored at the zero baseline, the ITM results ($M=1.2$) exhibit a higher variance, with price estimates fluctuating between $0.116$ and $0.167$ throughout the experimental window.

\begin{figure}[h]
    \centering
    \includegraphics[width=0.5\linewidth]{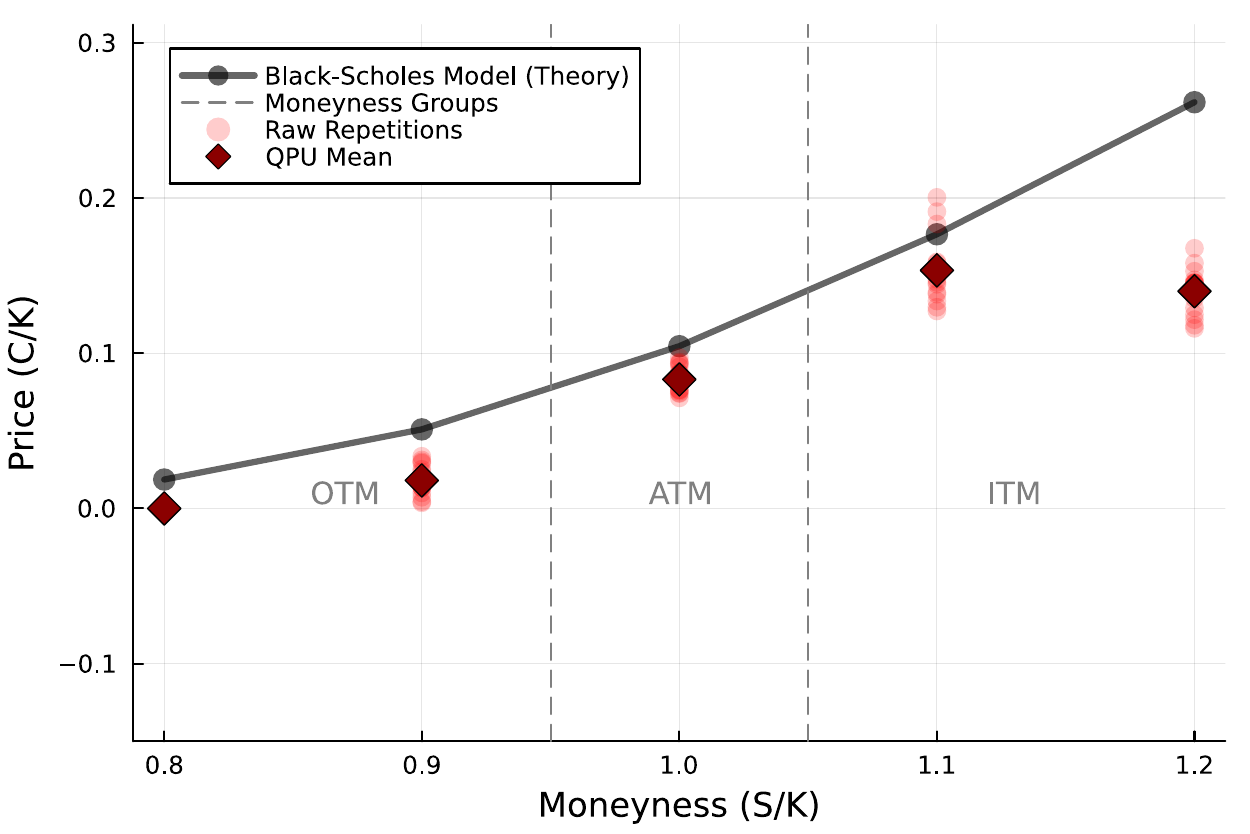}
    \caption{Point cloud distribution on Rigetti. A distinct systematic bias is visible as moneyness increases, while the ReLU transformation successfully anchors the OTM points to the zero-price baseline.}
    \label{fig:rigetti_cloud}
\end{figure}

The statistical distribution of these valuations is further illustrated in the point cloud analysis (Fig. \ref{fig:rigetti_cloud}). The data demonstrates a distinct "signal compression" effect, where a systematic negative bias becomes more pronounced as moneyness increases. This downward shift is most evident for the $M=1.2$ level, while the tight clustering of points at $M=1.0$ ($\sigma \approx 0.007$) confirms a high degree of operational repeatability for this architecture.

\subsubsection{Readout Error Mitigation and Accuracy Recovery}

To address the systematic negative bias observed in the Rigetti results, a Readout Error Mitigation (REM) protocol was implemented using a $2 \times 2$ Assignment Fidelity Matrix $A$. The calibration of the Ankaa-3 qubits yielded:

\begin{equation}
A = \begin{pmatrix} 0.97600 & 0.07260 \\ 0.02400 & 0.92740 \end{pmatrix}, \quad 
A^{-1} = \begin{pmatrix} 1.02657 & -0.08036 \\ -0.02657 & 1.08036 \end{pmatrix}
\end{equation}

By applying the inverse matrix $A^{-1}$ to the raw probability vectors, the model's global fit was significantly improved, as summarized in Table \ref{tab:mitigation_results}.

\begin{table}[H]
\centering
\caption{Global Accuracy Metrics for Rigetti Ankaa-3: Raw vs. Mitigated Data.}
\label{tab:mitigation_results}
\begin{tabular}{lccc}
\toprule
\textbf{Metric} & \textbf{Raw QPU} & \textbf{Mitigated QPU} & \textbf{Improvement (\%)} \\
\midrule
MSE             & 0.00345196 & \textbf{0.00252846} & -26.7\% \\
MAE             & 0.0435904  & \textbf{0.0345525}  & -20.7\% \\
RMSE            & 0.0587533  & \textbf{0.0502837}  & -14.4\% \\
$R^2$ Score     & 0.552264   & \textbf{0.672047}   & +21.7\% \\
\bottomrule
\end{tabular}
\end{table}

The error profile is non-uniform across the moneyness spectrum. OTM valuations ($m \le 0.8$) remain anchored to $0.0$ below the processing threshold, while peak performance is achieved for near-ATM options ($0.9 \le m \le 1.1$) with a mitigated $R^2$ of $0.8305$. In contrast, the $M=1.2$ level accounts for 85.94\% of the total MSE; although mitigation shifted the mean from $0.1399$ to $0.1560$, a significant gap relative to the theoretical $0.2617$ persists, indicating that in-circuit noise, rather than readout fidelity, is the limiting factor for deep ITM valuations.

%% file: 04_IBM.tex
\subsection{Experiments on the IBM Fez}

The hardware executions were conducted using the \texttt{ibm\_fez} quantum processor, accessed via the IBM Quantum Open Plan. This utility-scale platform provides a robust environment for validating QNN architectures, offering high-fidelity execution without the resource constraints typical of local simulators. To assess the stability and statistical convergence of the model, two distinct experimental configurations were implemented. The execution parameters and operational throughput are summarized in Table \ref{tab:qpu_params_ibm}.

\begin{table}[h]
\centering
\caption{Summary of IBM Fez Execution Parameters and Throughput.}
\label{tab:qpu_params_ibm}
\begin{tabular}{lcccc}
\toprule
\textbf{Experiment} & \textbf{Iterations ($R$)} & \textbf{Shots ($N$)} & \textbf{Total Circuits} & \textbf{Runtime} \\ \midrule
Dataset A (Standard) & 20 & 2,000 & 100 & 56 s \\
Dataset B (High-Fidelity) & 50 & 5,000 & 250 & 322 s \\ \bottomrule
\end{tabular}
\end{table}

\begin{figure}[htbp]
    \centering
    \begin{subfigure}[b]{0.48\textwidth}
        \centering
        \includegraphics[width=\textwidth]{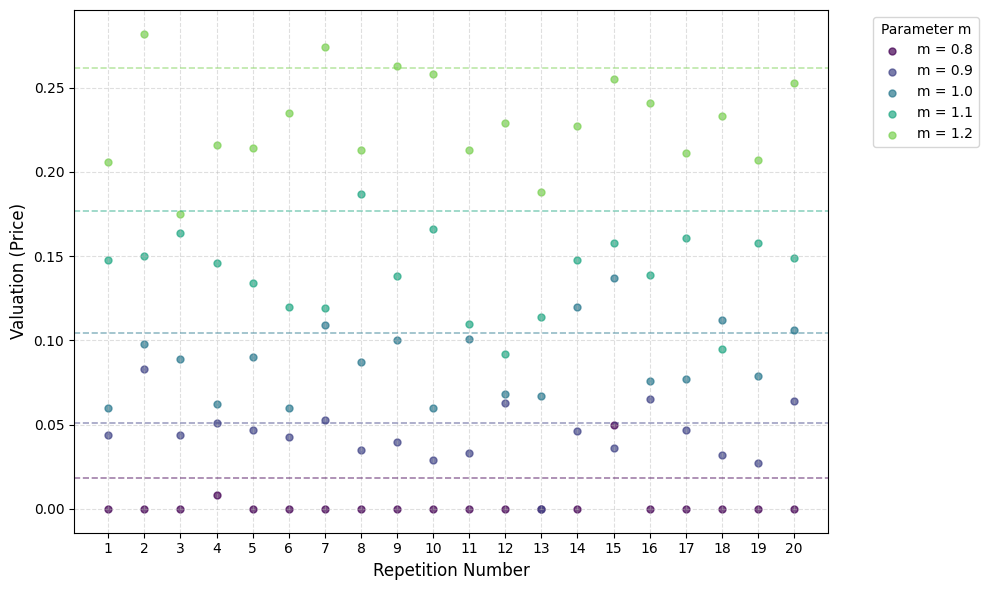}
        \caption{Dataset A: 20 iterations, 2k shots.}
        \label{fig:ibm_drift_a}
    \end{subfigure}
    \hfill
    \begin{subfigure}[b]{0.48\textwidth}
        \centering
        \includegraphics[width=\textwidth]{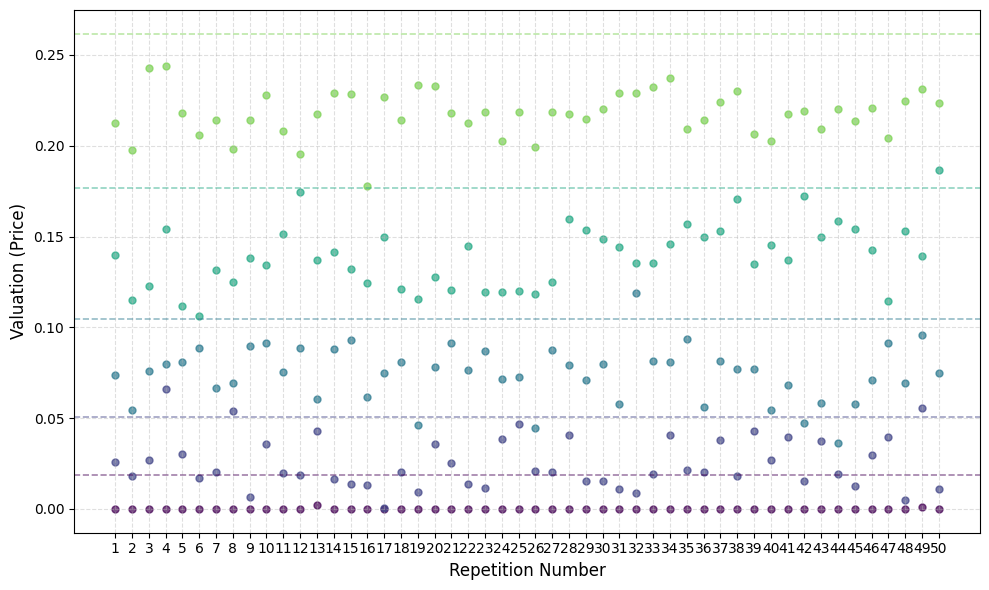}
        \caption{Dataset B: 50 iterations, 5k shots.}
        \label{fig:ibm_drift_b}
    \end{subfigure}
    \caption{Stability and drift analysis on \texttt{ibm\_fez}. Dashed lines represent the analytical Black-Scholes solution.}
    \label{fig:ibm_stability_drift}
\end{figure}

The system demonstrated high throughput, averaging less than 1.5 seconds per circuit for the high-fidelity configuration. Such efficiency is critical for the practical deployment of VQA and real-time derivative pricing. 
The stability of the QNN price estimator was monitored across multiple repetitions to identify potential temporal fluctuations, as shown in the drift analysis (Fig. \ref{fig:ibm_stability_drift}).

\begin{figure}[htbp]
    \centering
    \begin{subfigure}[b]{0.48\textwidth}
        \centering
        \includegraphics[width=\textwidth]{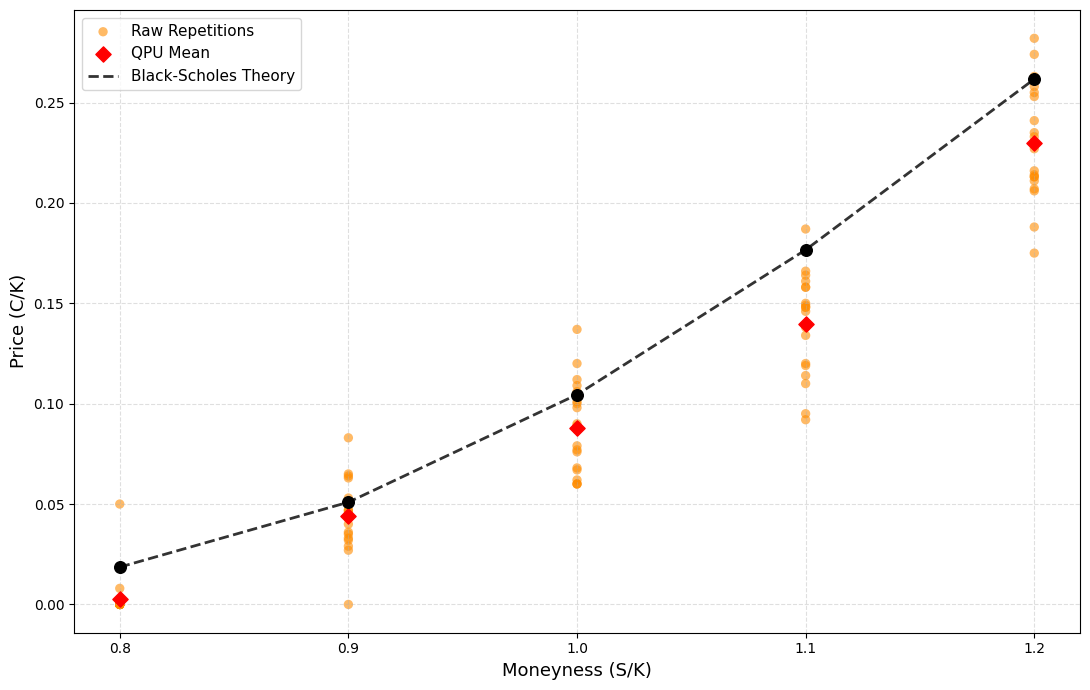}
        \caption{Dataset A (2k shots).}
        \label{fig:ibm_cloud_a}
    \end{subfigure}
    \hfill 
    \begin{subfigure}[b]{0.48\textwidth}
        \centering
        \includegraphics[width=\textwidth]{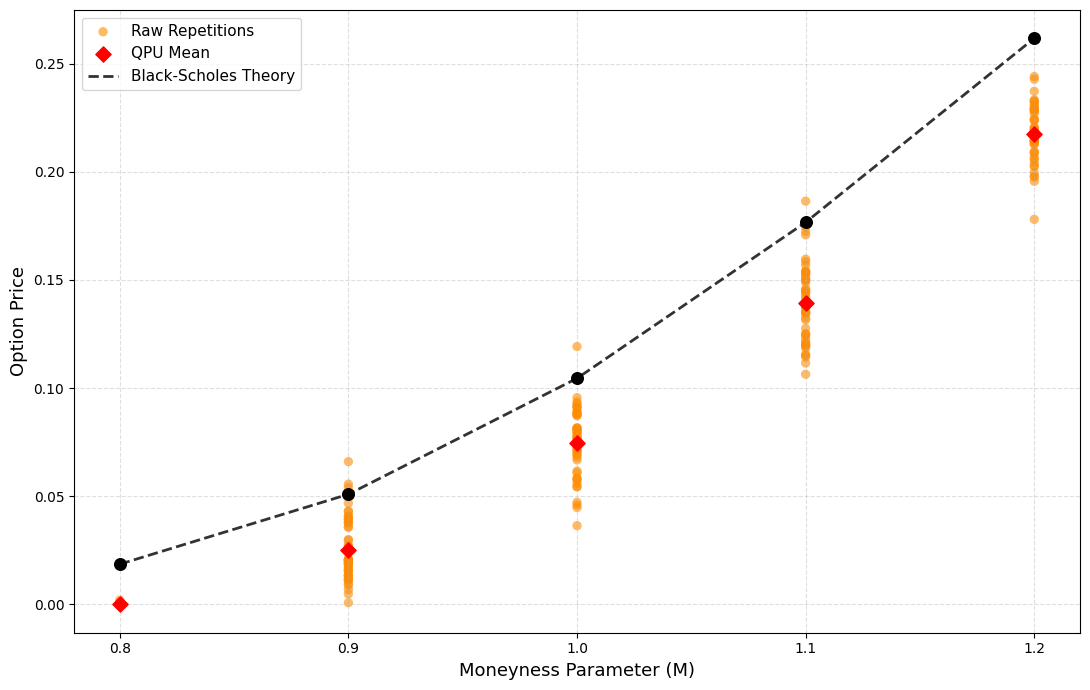}
        \caption{Dataset B (5k shots).}
        \label{fig:ibm_cloud_b}
    \end{subfigure}
    \caption{Point cloud distribution of valuations. The vertical shift reveals the hardware's systematic bias under different sampling depths.}
    \label{fig:ibm_cloud_analysis}
\end{figure}

The distribution of valuation samples is further illustrated in the point cloud analysis (Fig. \ref{fig:ibm_cloud_analysis}). 
The tight clustering of points indicates high repeatability and low device-level jitter. 
However, a consistent downward bias is observed in the In-the-Money (ITM) regions. 
This systematic attenuation suggests that as circuit complexity or amplitude requirements increase, hardware-induced decoherence shifts the estimates away from the theoretical baseline.

The experimental data reveals a critical accuracy-precision paradox on the \texttt{ibm\_fez} processor.
Despite the higher sampling volume, Dataset B achieved a lower $R^2$ score ($0.864$) compared to Dataset A ($0.923$). 
While 5,000 shots significantly increased the \textbf{precision} by reducing the statistical spread ($\sigma$), the extended execution time exposed the system to cumulative calibration drift. 
This temporal instability unmasked a systematic bias that was previously obscured by the larger statistical noise present in the 2,000-shot runs.
Consequently, these results suggest that for financial QNNs, increasing the shot count beyond a certain threshold may yield diminishing returns if the error budget becomes dominated by systematic hardware shifts rather than stochastic sampling noise.

\subsubsection{Ansatz Optimization and Unitary Decomposition}

To further mitigate the impact of hardware noise, the original QNN structure was compressed into a minimal hardware-efficient form using the method described in \ref{KAK_Decomposition}. This reduction significantly lowered the cumulative incoherent error and improved the fidelity of the state preparation. The optimized circuits were executed on \texttt{ibm\_fez} with a configuration of 40 repetitions and 4,000 shots per point ($2 \times 2,000$ to ensure temporal stability). 
The results of this compressed architecture are summarized in Table \ref{tab:ibm_cloud_results}.

\begin{table}[h]
\centering
\caption{Valuation Statistics for the Compressed 3-CX Ansatz on IBM Fez.}
\label{tab:ibm_cloud_results}
\begin{tabular}{ccccc}
\toprule
\textbf{M} & \textbf{BS Theory} & \textbf{QNN Mean} & \textbf{QNN Std} & \textbf{Bias} \\ \midrule
0.8 & 0.0186 & 0.0019 & 0.0045 & -0.0167 \\
0.9 & 0.0509 & 0.0272 & 0.0169 & -0.0237 \\
1.0 & 0.1045 & 0.0930 & 0.0165 & -0.0115 \\
1.1 & 0.1766 & 0.1590 & 0.0254 & -0.0177 \\
1.2 & 0.2617 & 0.2242 & 0.0199 & -0.0375 \\ \midrule
\multicolumn{5}{l}{\textbf{Global Metrics:} MSE: 0.000539, $R^2$: 0.9300} \\ \bottomrule
\end{tabular}
\end{table}
The compressed model achieved a superior $R^2$ score of 0.9300, outperforming all previous hardware configurations. This improvement is primarily associated with the reduced gate count, which minimized the accumulation of two-qubit gate errors. Despite the high global accuracy, a localized discrepancy was observed at $M = 0.8$, where valuations frequently approached zero. This effect occurs when the systematic downward bias shifts low-probability outcomes below the threshold of the post-processing transformation, resulting in a zero-price assignment. At the $M = 1.2$ level, a persistent systematic offset remains the dominant source of error, indicating that while circuit compression significantly improves fidelity, it does not entirely eliminate the bias inherent in the execution of the financial QNN architecture.

%% file: 05_results.tex
\section{Backend Synthesis and Comparative Analysis}

To evaluate the hardware-agnostic nature of the \textbf{finQbit} architecture, a comprehensive synthesis of results was conducted across four leading quantum computing platforms. This analysis aims to reconcile theoretical expectations with the physical constraints of the NISQ era, providing a benchmark for the model's portability. Table \ref{tab:global_qpu_comparison} summarizes the primary error metrics for each backend, utilizing the optimal sampling configuration identified in previous sections ($R \in \{20, 50\}$, $N=2,000$ shots).
\begin{table}[H]
\centering
\caption{Global Comparison of finQbit Performance across Physical QPUs.}
\label{tab:global_qpu_comparison}
\begin{tabular}{llcccc}
\toprule
\textbf{Backend} & \textbf{Technology} & \textbf{MAE} & \textbf{RMSE} & \textbf{MSE} & \textbf{$R^2$} \\ \midrule
\textit{Ideal Simulation} & \textit{State-Vector} & \textit{0.0152} & \textit{0.0163} & \textit{0.00027} & \textit{0.9654} \\ \midrule
\textbf{IBM Fez (U4)} & Superconducting & \textbf{0.0214} & \textbf{0.0232} & \textbf{0.00054} & \textbf{0.9300} \\
\textbf{IBM Fez (Std)} & Superconducting & 0.0216 & 0.0243 & 0.00059 & 0.9234 \\
\textbf{IonQ Forte} & Trapped Ion & 0.0482 & 0.0500 & 0.00250 & 0.6806 \\
\textbf{Rigetti Ankaa-3} & Superconducting & 0.0345 & 0.0503 & 0.00253 & 0.6720 \\
\textbf{IQM Garnet} & Superconducting & 0.0434 & 0.0519 & 0.00270 & 0.6495 \\ \bottomrule
\end{tabular}
\end{table}

The \textbf{IBM Fez} processor emerged as the most accurate physical platform, capturing the pricing surface with an $R^2$ exceeding $0.92$. This performance suggests that for low-qubit-count financial models, high-speed superconducting architectures with advanced runtime optimizations can closely approximate ideal state-vector simulations. 

The cross-platform analysis revealed distinct technology-specific bias signatures. Superconducting systems (IBM, Rigetti, IQM) consistently exhibited a systematic downward price bias in the ITM region, whereas the trapped ion architecture (\textbf{IonQ}) demonstrated higher temporal stability but maintained a persistent upward shift. These divergent profiles suggest that while superconducting qubits are limited by signal attenuation, trapped ion systems are primarily influenced by systematic gate calibration offsets. Despite these varying noise signatures and native gate sets, the \textbf{finQbit} model maintained a predictive $R^2 > 0.64$ across all tested hardware. This architectural robustness confirms that the model's inductive bias, rooted in Hilbert space rotations, remains effective against diverse decoherence profiles, establishing its viability for multi-platform financial deployments.

\section{Conclusion and Outlook}

This work has presented a Quantum Neural Network-based framework for option pricing on Noisy Intermediate-Scale Quantum devices. 
Using the Black-Scholes-Merton model as a controlled benchmark, we investigated whether compact variational quantum circuits can learn and reproduce the option pricing surface with sufficient accuracy to be meaningful in a financial context. 
The use of the BSM setting was deliberate. It provides a well-understood analytical reference, thereby allowing the predictive capabilities of the quantum model to be assessed rigorously and without ambiguity.

To the best of our knowledge, this study constitutes one of the first end-to-end implementations of a fully quantum option pricing workflow executed on currently available quantum hardware. 
In contrast to hybrid schemes in which the quantum component is limited to an isolated subroutine, the present approach uses a Quantum Neural Network as the core pricing engine and validates it across multiple physical backends, including IBM Fez, IQM Garnet, IonQ Forte, and Rigetti Ankaa-3. 
This cross-platform execution is a central contribution of the paper, as it moves the discussion beyond idealized simulation and into the domain of empirical hardware benchmarking.

The numerical results demonstrate that accurate option pricing approximations can already be achieved on real quantum processors despite the constraints of the NISQ era. 
Most notably, the proposed \textit{finQbit} architecture attains predictive performance comparable to strong classical machine learning baselines while requiring a remarkably compact circuit and a relatively small number of trainable parameters. 
The empirical analysis further shows that the architecture is robust across heterogeneous hardware technologies and noise profiles, although the detailed bias characteristics vary substantially by platform. 
These findings provide concrete evidence that quantum machine learning can serve as a viable approximation framework for financial pricing functions under present-day hardware limitations.

Beyond the numerical results themselves, this work highlights several conceptual insights. 
First, the continuous geometry of Hilbert space appears well-suited to the representation of smooth financial manifolds such as the Black-Scholes-Merton pricing surface. 
Second, careful architectural compression, including feature-dense encodings and unitary consolidation, is essential for practical deployment on noisy hardware. 
Third, the observed trade-off between estimator variance, shot count, and execution cost confirms that benchmarking quantum financial models must be conducted not only in terms of predictive accuracy, but also in terms of operational feasibility.

At the same time, the present study should be understood as a proof of concept rather than a final application-level solution. 
The Black-Scholes-Merton model is intentionally simple relative to the models used in production settings, where practitioners routinely face stochastic volatility, local volatility, interest rate dynamics, jumps, and path dependence. 
The true significance of the present results, therefore, lies not in replacing classical BSM pricing but in establishing that QNN-based methods can already function reliably on real hardware and may be extended to settings where classical methods become computationally burdensome.

Several natural directions for future research follow from this work. 
A first priority is the extension of the framework to more realistic stochastic models, including Heston, local volatility, and SABR dynamics. 
Such models introduce richer nonlinear structure and would provide a more demanding test of the expressive power of Quantum Neural Networks. 
A second direction concerns the treatment of exotic and path-dependent derivatives, where the advantage of learned surrogate models may be more pronounced than in closed-form settings. 
This includes products such as Asian options, barrier options, and American-style contracts, for which valuation typically requires costly numerical procedures. 
A third direction involves the integration of more advanced quantum estimation techniques, including Quantum Amplitude Estimation–type methods, in order to reduce the sampling burden that currently limits hardware execution efficiency. 
Finally, continued work on ansatz design, error mitigation, and backend-aware transpilation will be essential for narrowing the gap between ideal simulations and physical device performance.

In summary, the results presented here suggest that Quantum Neural Networks are not merely a speculative tool for distant fault-tolerant machines, but a practically meaningful modeling framework already worth studying in the NISQ regime. 
While substantial hardware and algorithmic advances are still required, the present work indicates that the foundations for quantum-enhanced derivative pricing are beginning to take shape. 
As quantum hardware matures, methods of the type developed here may form part of a broader computational toolkit for financial engineering, particularly in applications where classical methods face severe complexity, latency, or scalability constraints.